\documentclass[showpacs,aps,twocolumn]{revtex4}
\usepackage{epsfig}
\usepackage{graphicx}
\usepackage{amsmath,amssymb,amsfonts}
\usepackage{amsmath}
\usepackage{array}
\usepackage{url}
\usepackage{multirow}
\usepackage{float}
\usepackage{comment}

\usepackage{booktabs}  
\usepackage{longtable} 

\usepackage{lineno}
\usepackage{xspace}
\usepackage{ulem}

\newcommand{\bef}{\begin{figure}}
\newcommand{\eef}{\end{figure}}
\newcommand{\bc}{\begin{center}}
\newcommand{\ec}{\end{center}}

\newcommand{\be}{\begin{equation}}
\newcommand{\ee}{\end{equation}}
\newcommand{\bea}{\begin{eqnarray}}
\newcommand{\eea}{\end{eqnarray}}

\def\ba{\begin{eqnarray}}
\def\ea{\end{eqnarray}}

\usepackage[usenames,dvipsnames]{color}
\definecolor{darkblue}{RGB}{0,0,196}
\usepackage[colorlinks=true,linkcolor=darkblue,citecolor=darkblue,urlcolor=darkblue]{hyperref}

\begin{document}
\title{Einstein–de Haas effect and induced rotation in QCD matter}

\author{Dushmanta Sahu}
\email{Dushmanta.Sahu@cern.ch}
\affiliation{Instituto de Ciencias Nucleares, UNAM, Apartado Postal 70-543, Coyoacán, 04510, México City, México}

\begin{abstract}
In this study, we report the first identification of the Einstein–de Haas (EdH) effect in the QCD matter. The EdH effect is a fundamental magnetomechanical coupling wherein magnetic-field-induced spin alignment generates a compensating collective rotation to conserve the total angular momentum. Using an equilibrium hadron gas under an external magnetic field, we show that even remnant magnetic fields at the freeze-out produce induced rotations ($\omega_{\mathrm{EdH}}$) comparable to typical estimates of fluid vorticity in heavy-ion collisions as inferred from final-state hyperon polarization. This rotation emerges from the magnetic field alone, without any initial vorticity as input. The Einstein–de Haas effect thus establishes hot QCD matter as a self-vortical magnetofluid, where collective rotation can be generated purely from spin alignment, and identifies spin-rotation coupling as a potentially important, previously overlooked component of angular momentum dynamics in relativistic nuclear collisions.
\end{abstract}

\maketitle

\textit{Introduction—} Relativistic heavy-ion collisions generate the most vortical fluid in the laboratory. In non-central heavy-ion collisions, the large orbital angular momentum of the participants converts into fluid vorticity, which in turn polarizes hadron spins through spin-vorticity coupling. Measurements of $\Lambda$ hyperon polarization at RHIC have confirmed this picture~\cite{STAR:2017ckg}, establishing a direct connection between collective motion and spin degrees of freedom in quantum chromodynamics (QCD) matter. On the other hand, relativistic peripheral heavy-ion collisions are also known to produce a huge magnetic field (of the order of $m_{\pi}^2 \sim 10^{18}$ G) due to the motion of the spectator protons. This magnetic field is transient, decays with time, and in principle affects the thermodynamic and transport properties of the evolving partonic and hadronic matter~\cite{Skokov:2009qp,Yan:2021zjc}. These two unique consequences of heavy-ion collisions have fundamentally opened new avenues of research regarding phase diagram of QCD matter~\cite{Pradhan:2021vtp,Sahoo:2023vkw,Goswami:2023eol,Pradhan:2023rvf} and other interesting phenomena, such as chiral vortical effects, chiral magnetic effects~\cite{Kharzeev:2010gr,Kharzeev:2015znc}.

Recently, studies have shown how a rotating medium can generate a finite magnetic field comparable to the external magnetic fields at RHIC energies due to the Barnett effect~\cite{Sahu:2025tmb}. In condensed-matter and atomic systems, the Einstein–de Haas (EdH) effect is the reciprocal of Barnett effect, when a magnetic field aligns spins, and then conservation of total angular momentum demands a compensating mechanical rotation. The EdH effect, first demonstrated in 1915~\cite{EinsteinDeHaas1915}, is a phenomenon in which a ferromagnetic cylinder suspended by a torsion fiber begins to rotate when its magnetization is altered. Recently, this has also been explored in quantum systems by considering a single-molecule magnet coupled to a nanomechanical resonator~\cite{Ganzhorn2016}. The phenomenon is also explored in an optically confined Bose-Einstein condensate of europium atoms, where after the atoms were transferred from the highest to lower spin states, the exchange of angular momentum between atomic spins and the rotation of the quantum fluid is observed~\cite{Matsui}. This magnetomechanical coupling is a consequence of angular momentum conservation. While the Barnett effect in QCD matter has been recently explored~\cite{Sahu:2025tmb}, the EdH response has not yet been investigated in the context of heavy-ion collisions. Together, these effects establish a dynamical coupling between spin and orbital angular momentum in strongly interacting matter.

The hadronic phase of the fireball provides an ideal environment for the EdH effect. It contains a rich spectrum of particles with large magnetic moments, persists for $\sim 5 \-- 10~\mathrm{fm}/c$, and experiences residual magnetic fields. Moreover, the temperature near the transition temperature of quark-gluon plasma maximizes the magnetic susceptibility of the hadron gas. If these magnetic fields align hadronic spins, creating net spin angular momentum $S_z$, then the angular momentum conservation forces the system to develop a compensating rotation ($\omega_{\mathrm{EdH}}$) satisfying $\Delta L + \Delta S = 0$.

The development of relativistic spin hydrodynamics and kinetic approaches, in which spin degrees of freedom are incorporated into the fluid description and evolve under the influence of vorticity and electromagnetic fields, has led to significant advances in our understanding of QCD matter in heavy-ion collisions~\cite{Florkowski:2018fap,Singh:2022ltu}. Within these frameworks, the dominant paradigm is that fluid vorticity induces spin polarization through spin–vorticity coupling, with polarization evaluated either via thermal vorticity at freeze-out or through dynamical transport evolution. Despite substantial progress, including the formulation of spin transport equations and detailed studies of polarization dynamics~\cite{Gao:2012ix,Weickgenannt:2019dks}, several challenges remain, such as the sensitivity to pseudogauge choices and the unresolved longitudinal polarization sign problem~\cite{Karpenko:2016jyx,STAR:2019erd}. Parallel developments in anomalous transport have explored the role of vorticity and magnetic fields in generating axial and vector currents through the chiral vortical and chiral magnetic effects, further highlighting the intricate interplay between rotation, spin, and electromagnetic fields in QCD matter~\cite{Fukushima:2008xe,Son:2012wh}. However, existing approaches predominantly treat spin polarization as a response to vorticity or external electromagnetic fields, without explicitly accounting for the reciprocal backreaction of spin alignment on the macroscopic angular momentum of the fluid. While angular momentum conservation underlies spin hydrodynamics, its realization as a dynamical conversion between spin and orbital components has not been quantitatively established in the context of heavy-ion collisions. 

In this work, we provide the first quantitative investigation of the EdH effect in a hot hadronic medium. Using a hadron gas under external magnetic field, with Landau quantization and magnetic moments, we estimate the magnetic-field-induced spin polarization of all hadronic species and the resulting backreaction on the fluid's angular velocity through angular-momentum conservation. We find that even remnant magnetic fields at freeze-out produce induced rotation $\omega_{\mathrm{EdH}}$ comparable to typical estimates of initial fluid vorticity. Remarkably, this rotation emerges solely from the magnetic field, without requiring initial vorticity. Consequently, the vorticity inferred from spin polarization may differ from the hydrodynamic vorticity, but a net vorticity that may be substantially reduced by magnetomechanical backreaction. The Einstein–de Haas effect thus establishes hot QCD matter as a self-vortical magnetofluid, where spin and orbital angular momentum are continuously coupled throughout the fireball evolution. This highlights a previously overlooked component of angular momentum dynamics that provides a quantitatively relevant correction under realistic conditions in any complete description of spin phenomena in heavy-ion collisions.
\\

\begin{figure*}
    \centering
    \includegraphics[width=0.8\linewidth]{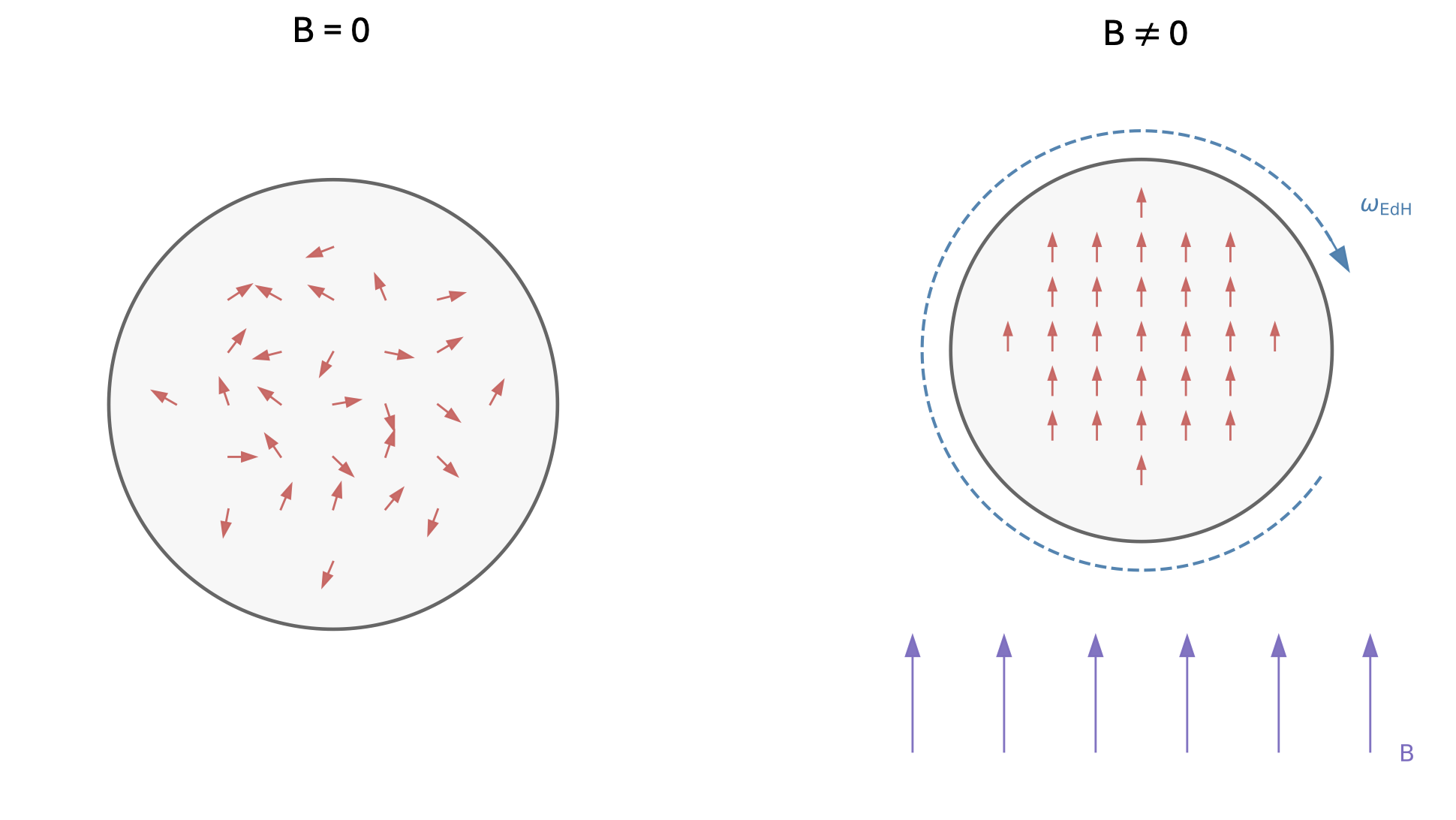}
    \caption{A schematic representation of how magnetic field creates a spin alignment, which in turn creates a rotation to conserve the total angular momentum.}
    \label{fig1}
\end{figure*}

\textit{Einstein--de Haas effect in hadron gas—}We describe the hadronic phase as a non-interacting gas of hadrons in global equilibrium at temperature $T$ and baryon chemical potential $\mu_B$ in the presence of an external magnetic field $B$. The list of hadrons used in this work is taken from table~\ref{tab:hrg_particles}. The total pressure of the system is defined as,

\begin{equation}
P(T,\mu_B,B) = \sum_i P_i(T,\mu_i,B), \nonumber
\end{equation} where
\begin{equation}
P_i(T,\mu_i, B) = T d_i \int \frac{d^3 p}{(2\pi)^3} \ln \left[ 1 \pm \exp\left(-\frac{E_i - \mu_i}{T}\right) \right] 
\label{Eq1}
\end{equation}

where the single-particle energies are modified by the field. Here, $p$ is the momentum and $d_i$, $m_i$, $E_i$, $\mu_{i}$ are the degeneracy, mass, energy and the total chemical potential of the $i$-th particle respectively. For our study, we consider only the baryon chemical potential, $\mu_{B}$. The $\pm$ signs are used for fermions ($+$) and bosons ($-$) in the system.

For charged hadrons, Landau quantization and magnetic-moment interaction gives,

\begin{equation}
E_{c,i}(p_z,k,s_z) = \sqrt{p_z^2 + m_i^2 + 2|q_i|eB\left(k + \tfrac{1}{2}\right)} - \mu_i^* B s_z, \nonumber
\end{equation}

where $p_{z}$ and $s_{z}$ are the $z$-component of the momentum and spin, and $k = 0,1,2,\dots$ is the Landau index. $q_i$ and $\mu_i^*$ are the particle charge and the magnetic moment ($\mu_i^* = \frac{g_i e}{2 m_i}$) of the $i$-th particle, $g_i$ is the Land\'e $g$-factor of the particle, whereas $e$ is the elementary charge. In the presence of Landau levels, the three-dimensional momentum integral can be reduced to a one-dimensional integral~\cite{Fraga:2008qn}:
\begin{equation}
\int \frac{d^3 p}{(2\pi)^3}
=
\frac{|q_i| e B}{2\pi}
\sum_{k} \alpha_k \int \frac{dp_z}{2\pi},\nonumber
\end{equation}
where $\alpha_k = 2 - \delta_{k0}$.

Neutral hadrons experience only the Zeeman term,

\begin{equation}
E_{n,i}(p,s_z) = \sqrt{p^2 + m_i^2} - \mu_i^* B s_z. \nonumber
\end{equation}

The magnetization follows from the thermodynamic relation,

\begin{equation}
\mathcal{M}_z(T,\mu_B,B) = \left(\frac{\partial P}{\partial B}\right)_{T,\mu_B} = \sum_i \left(\frac{\partial P_i}{\partial B}\right)_{T,\mu_B}. 
\label{Eq2}
\end{equation}

This gives the net magnetic moment per unit volume. However, the quantity required for angular momentum conservation is the spin angular momentum density, $\mathcal{S}_z$. To connect these, recall that for a particle of species $i$, the magnetic moment operator $\hat{\mu}^{*}_z$ is proportional to the spin operator $\hat{s}_z$,
\begin{equation}
\hat{\mu}^{*}_z = g_i \frac{e}{2m_i} \hat{s}_z.\nonumber
\end{equation}
Taking the expectation value in a thermal state gives $\langle \hat{\mu}^{*}_z \rangle_i = \frac{g_i e}{2m_i} \langle \hat{s}_z \rangle_i$. The total magnetization density $\mathcal{M}_z = \sum_i n_i \langle \hat{\mu}^{*}_z \rangle_i$ is related to the pressure via Eq.~\ref{Eq2}, while the spin angular momentum density is $\mathcal{S}_z = \sum_i n_i \langle \hat{s}_z \rangle_i$. Combining these relations yields,
\begin{equation}
\mathcal{S}_z(T,\mu_B,B) = \sum_i \frac{2m_i}{g_i e} \left(\frac{\partial P_i}{\partial B}\right)_{T,\mu_B}. 
\label{Eq3}
\end{equation}

This is the thermodynamic expression for the net spin alignment that must be compensated by mechanical rotation. 

At this point, it is important to distinguish between spin and orbital contributions to the magnetization. Landau quantization generates an orbital magnetization associated with microscopic cyclotron motion, corresponding to internal circulating currents. In contrast, the Einstein--de Haas effect involves the conversion of intrinsic spin angular momentum into a macroscopic rigid-body rotation, $L_{\mathrm{mech}} = I \omega$. Therefore, only the spin component can drive a global rotation of the fluid, while the orbital contribution from Landau levels remains confined to internal motion and does not induce a net torque. Accordingly, in evaluating the angular momentum balance relevant for the Einstein--de Haas response, we retain only the spin contribution from the Zeeman coupling, while Landau-level effects are included solely in the thermodynamics. We note that the decomposition of total angular momentum into spin and orbital parts is not unique in relativistic quantum field theory and depends on the choice of pseudogauge~\cite{Becattini:2018duy}. Here we adopt the canonical formulation~\cite{Groot,Becattini:2013fla}. While different choices may affect the local separation, the total angular momentum conservation ensures that the emergence of a compensating collective rotation from spin alignment remains a physically robust feature.

A rotating medium induces a spin polarization through the coupling of angular velocity $\boldsymbol{\omega}$ to spin, with energy shift $\Delta E_{\mathrm{rot}} = -\boldsymbol{\omega} \cdot \mathbf{s}$ (in units where $\hbar=1$). This is directly analogous to the Zeeman shift $\Delta E_{\mathrm{mag}} = -\boldsymbol{\mu^{*}} \cdot \mathbf{B}$ in a magnetic field. For a given particle species, equating the two shifts gives, $-\omega s_z = -\mu^{*}_z B$, assuming rotation along the z-axis, $\boldsymbol{\omega} = \omega \hat{z}$. Substituting $\mu^{*}_z = \frac{g_i e}{2m_i} s_z$ from the gyromagnetic relation, the spin quantum number $s_z$ cancels (provided the particle is polarized, i.e., $\langle s_z \rangle \neq 0$), leading to,
\begin{equation}
\omega = \frac{g_i e}{2m_i} B \quad \Longrightarrow \quad B = \frac{2m_i}{g_i e} \omega. 
\label{Eq4}
\end{equation}
This relation holds at the level of a given particle species; for a multi-component system, it should be interpreted as an effective mapping between rotation and magnetic response. Physically, Eq.~\ref{Eq4} states that a rotating frame is equivalent, for spin degrees of freedom, to a magnetic field of magnitude $B$, proportional to the angular velocity. This equivalence underlies the Einstein--de Haas effect: a magnetic field aligns spins, which is equivalent to imparting a mechanical rotation, and vice versa. The Einstein--de Haas effect enforces conservation of total angular momentum,

\begin{equation}
L_{\mathrm{tot}} = L_{\mathrm{mech}} + L_{\mathrm{spin}} = \mathrm{const}. \nonumber
\end{equation}

Because of this, an initially non-rotating system develops a mechanical rotation when spins align. The total spin angular momentum is,

\begin{equation}
L_{\mathrm{spin}} = V \mathcal{S}_z, 
\label{Eq5}
\end{equation}

and the mechanical part is,

\begin{equation}
L_{\mathrm{mech}} = I \omega_{\mathrm{EdH}}, \nonumber
\end{equation}

where $V$ is the volume of the system and $I$ is the moment of inertia of the fireball. For a cylindrical system of radius $R$ and energy density $\epsilon$,

\begin{equation}
I = \frac{1}{2} \epsilon V R^2. 
\label{Eq6}
\end{equation}

The moment of inertia is estimated assuming a rigidly rotating cylindrical fireball with uniform energy density, as in a relativistic system, the inertia is governed by energy density rather than mass density. The single particle energy density is given as,

\[
\epsilon_i(T,\mu_i, B)
=
d_i \int \frac{d^3 p}{(2\pi)^3}
\frac{E_i}{\exp\left(\frac{E_i - \mu_i}{T}\right) \pm 1},
\]
and \[
\epsilon(T,\mu_B, B) = \sum_i \epsilon_i(T,\mu_i, B).
\]

In realistic heavy-ion collisions, however, the system exhibits strong longitudinal expansion. More generally, the moment of inertia can be given by
\begin{equation}
I = \int d^3x \, \epsilon(\mathbf{x}) \, r_\perp^2, \nonumber
\end{equation}
which can be expressed parametrically as
\begin{equation}
I \sim \kappa \, \epsilon V R^2, \nonumber
\end{equation}
where $\kappa = \mathcal{O}(1)$ encodes the effects of the spatial energy density distribution. For typical profiles relevant to heavy-ion collisions (e.g., Gaussian or Glauber initial conditions), one expects $\kappa \sim 0.3\text{--}1$. This introduces an $\mathcal{O}(1)$ uncertainty in the magnitude of the induced angular velocity, but does not affect its parametric dependence on thermodynamic quantities. Consequently, our estimates should be interpreted as order-of-magnitude predictions that are robust against geometric details of the fireball.

Conservation of total angular momentum, $I\omega_{\mathrm{EdH}} + V \mathcal{S}_z = 0$, then gives the induced rotation,

\begin{equation}
\omega_{\mathrm{EdH}}(T,\mu_B,B) = -\frac{2 \mathcal{S}_z}{\epsilon R^2}. 
\label{Eq7}
\end{equation}

This induced rotation acts in opposition to the magnetic-field--driven spin alignment (signified by the negative sign), leading to a net suppression of the final vorticity and subsequently a suppression of the observable hyperon polarization. Crucially, $\omega_{\mathrm{EdH}}$ depends only on thermodynamic quantities and fireball geometry, making it a robust, model-independent prediction of angular momentum conservation in magnetized matter. The relation $I \omega_{\mathrm{EdH}} + V S_z = 0$ implicitly assumes that the spin angular momentum generated by the magnetic field is converted into a coherent, global rotation of the medium. In a realistic heavy-ion collision, however, angular momentum can also be redistributed into local vorticity, shear flow, or internal excitations. The rigid-rotation limit adopted here therefore corresponds to a maximal collective response, in which the entire spin angular momentum is transferred to the lowest-mode orbital motion of the system. Any redistribution into non-collective or higher-order flow modes would reduce the net global angular velocity. Consequently, Eq.~\ref{Eq7} should be interpreted as providing an upper bound on the magnitude of the Einstein--de Haas--induced rotation, with more realistic dynamical evolution expected to yield a smaller effect while preserving the same parametric dependence on thermodynamic quantities.

Several simplifications underlie our calculation. First, we assume global equilibrium with a uniform magnetic field, neglecting spatial gradients and time evolution; the cumulative memory of spin alignment over the fireball lifetime ensures that our equilibrium estimate captures the leading-order magnetization. Second, we model the fireball as a rigidly rotating cylinder with moment of inertia $I = \frac{1}{2} \epsilon V R^2$; more realistic cases will modify the geometric prefactor by $\mathcal{O}(1)$ and do not alter the parametric scaling. Third, the induced rotation $\omega_{\mathrm{EdH}}$ is the mechanical angular velocity; for rigid rotation, the kinetic vorticity is $2\omega_{\mathrm{EdH}}$, a factor that lies within the uncertainties of our comparisons. Fourth, neutral hadron magnetic polarizabilities and non-equilibrium corrections are sub-leading effects that will be addressed in future work. None of these assumptions affect our central conclusion that angular momentum conservation forces a collective rotation of magnitude that can reach the same order of magnitude as typical vorticity estimates, establishing the Einstein--de Haas effect as an essential component of spin dynamics in heavy-ion collisions.
\\

\begin{figure}[t]
    \centering
    \includegraphics[width=0.99\linewidth]{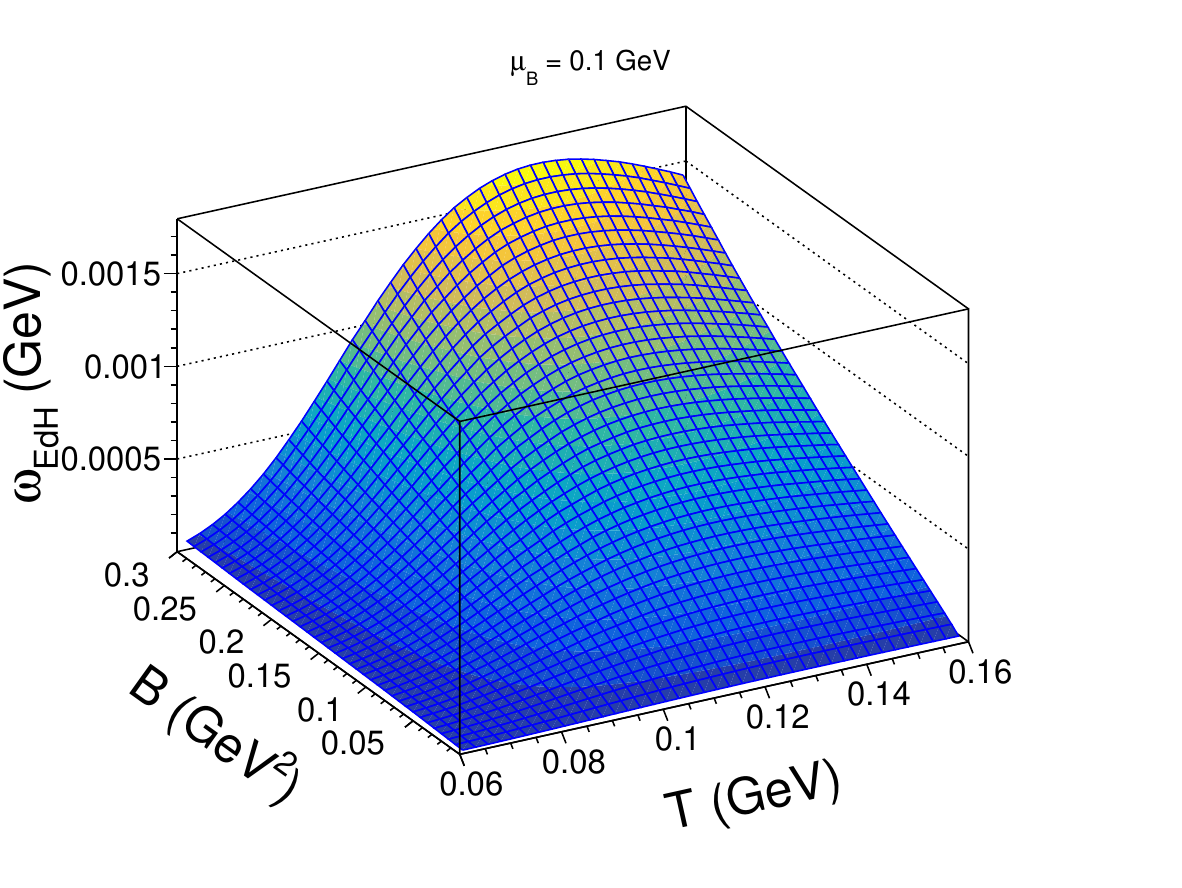}
    \caption{Induced rotation (shown as absolute value) as a function of temperature and external magnetic field for a fixed baryochemical potential $\mu_{B} = 0.1$ GeV.}
    \label{fig2}
\end{figure}

\textit{Results and Discussion—} In heavy-ion collisions, the primary magnetic field from spectator protons decays rapidly, with a typical timescale $\tau_B \sim 1$--$2$~fm/$c$ at RHIC energies. However, the conducting QCD medium responds via Faraday induction, where a time-varying magnetic field generates an electric field, which drives currents that partially sustain the magnetic field~\cite{Shen:2025unr,Tuchin:2013apa}. Our analysis therefore considers the range $B = 10^{-4}$--$10^{-1}$~GeV$^2$, with the lower end representing realistic freeze-out remnants and the upper end providing an upper bound. 

Figure~\ref{fig1} illustrates the Einstein--de Haas mechanism in a hadronic fireball with zero initial rotation. In the absence of a magnetic field, hadron spins are randomly oriented, as shown in the left panel. When a magnetic field is applied, spins align preferentially along the field direction, generating a net spin angular momentum. Conservation of total angular momentum then requires the system to develop a compensating collective rotation $\omega = \omega_{\mathrm{EdH}}$ opposite to the spin alignment. This backreaction is entirely determined by the magnetization of the hadronic medium.

Figure~\ref{fig2} shows the induced rotation $\omega_{\mathrm{EdH}}$ as a function of temperature and magnetic field at a fixed baryochemical potential of $\mu_{B}$ = 0.1 GeV. At low fields ($B \sim 10^{-4}$ GeV$^2$), $\omega_{\mathrm{EdH}}$ increases approximately linearly with $B$, reflecting the growth of spin polarization. At high temperatures, $\omega_{\mathrm{EdH}}$ decreases slightly because of the thermal motion diluting the spin alignment, which causes a reduction in $\omega_{\mathrm{EdH}}$. However, one can observe that the induced rotation is substantial and in some cases comparable to the vorticity estimated from final state hyperon polarization, which is almost negligible at ALICE energies, while at lower RHIC energies it reaches values of order $\sim 0.01$--$0.02$ GeV at most~\cite{STAR:2017ckg,ALICE:2019onw}. This induced rotation acts against the initial vorticity and hence reduces the final net vorticity. These results demonstrate that magnetomechanical backreaction can significantly modify the effective vorticity of the system. The vorticity inferred from final-state polarization therefore reflects a net vorticity,
\begin{equation}
\omega_{\mathrm{total}} = \omega_{\mathrm{hydro}} + \omega_{\mathrm{EdH}}, \nonumber
\end{equation}

which may be substantially reduced by the EdH contribution. In this sense, spin alignment induced by magnetic fields feeds back into the orbital motion of the fluid, establishing hot QCD matter as a self-vortical magnetofluid in which magnetic field, spin polarization, and rotation are dynamically coupled. 

Realistically, the magnitude of this effect depends on the interplay between the magnetic-field decay and the spin relaxation dynamics. The spin polarization evolves toward its equilibrium value over a characteristic relaxation time $\tau_s$, which can be modeled by a relaxation-time equation~\cite{Bhadury:2020puc},
\begin{equation}
\frac{dS_z(t)}{dt} = \frac{S_z^{\mathrm{eq}}(T,B(t)) - S_z(t)}{\tau_s}, \nonumber
\end{equation}
with $S_z^{\mathrm{eq}} \propto \chi_s\,B(t)$, where $\chi_{s}$ is the spin susceptibility. Assuming an exponentially decaying magnetic field $B(t) = B_0 e^{-t/\tau_B}$~\cite{Deng:2012pc}, the accumulated spin polarization, and hence the induced rotation, depends on the competition between $\tau_s$ and the magnetic-field decay time $\tau_B$. Parametrically, this leads to
\begin{equation}
\omega_{\mathrm{EdH}} \sim \omega_{\mathrm{eq}} \frac{\tau_B}{\tau_B + \tau_s},\nonumber
\end{equation}
where $\omega_{\mathrm{eq}}$ denotes the equilibrium estimate. In the fast relaxation limit ($\tau_s \ll \tau_B$), the system approaches the equilibrium result, while for $\tau_s \gtrsim \tau_B$, the induced rotation is suppressed. Since estimates of $\tau_s$ being a few fm$/c$ are comparable to the fireball lifetime~\cite{Ayala:2019iin}, partial memory of early-time magnetic fields may survive and contribute to the observed rotation~\cite{Tuchin:2013apa}. This indicates that the equilibrium HRG calculation provides an upper bound on the magnitude of the EdH-induced effect under realistic conditions.

\textit{Conclusion—} We have identified the Einstein--de Haas effect as a fundamental magnetomechanical coupling in hot QCD matter. In the hadronic phase, magnetic-field-induced spin alignment generates a compensating collective rotation through angular momentum conservation. Using a hadron gas including Landau quantization and magnetic moments, we find that even remnant magnetic fields at freeze-out can induce rotation comparable to typical estimates of fluid vorticity at RHIC and the LHC energies. This rotation can arise solely from the magnetic field, even in the absence of initial vorticity. As a result, the vorticity inferred from spin polarization reflects a net vorticity modified by magnetomechanical backreaction, rather than the initial vorticity generated in the collision. The EdH effect establishes hot QCD matter as a self-vortical magnetofluid, in which spin and orbital angular momentum are dynamically coupled. Although the magnetic field is short-lived and the system is not in strict equilibrium, the EdH mechanism follows directly from angular momentum conservation and remains operative for residual fields. These results indicate that any mechanism generating spin polarization through magnetic fields must be accompanied by a compensating orbital rotation. Together with the Barnett effect, this points to a coupled spin--vorticity dynamics in QCD matter, where spin and orbital angular momentum continuously interconvert, necessitating its inclusion in any complete description of spin phenomena in heavy-ion collisions. However, a full dynamical picture is still needed in order to properly understand the effect of EdH on the final state polarization of hyperons or vector mesons. Moreover, a quantitative estimate of the resulting modification to hyperon polarization can be explored in future studies.

\section*{Acknowledgement}
DS acknowledges the support from the postdoctoral fellowship of the DGAPA UNAM.
\\

\section*{Appendix}
\begin{table*}
\centering
\caption{The complete hadron gas particle list used in this study (particles Only)~\cite{ParticleDataGroup:2024cfk}. For every particle listed below, there exists a corresponding anti-particle with identical mass, spin, and degeneracy, but with opposite charge, opposite baryon number, and opposite magnetic moment. For neutral particles that are their own anti-particles (e.g., $\pi^0$, $\eta$, etc.), the properties listed apply to the particle itself. For some of the vector mesons, the g-factor is taken from Ref.~\cite{QCDSF:2008tjq,Badalian:2012ft,Luschevskaya:2018sbp}.}
\label{tab:hrg_particles}
\footnotesize
\begin{tabular}{|@{\extracolsep{\fill}}c|c|c|c|c|c|c@{}|}
\hline
Name & Mass (GeV) & Charge ($e$) & Spin & Degeneracy & g-factor & Baryon Number \\
\hline
$\pi^0$ & 0.13498 & 0 & 0.0 & 1 & 0.0 & 0 \\
\hline
$\pi^+$ & 0.13957 & 1 & 0.0 & 1 & 0.0 & 0 \\
\hline
$K^0_L$ & 0.49761 & 0 & 0.0 & 1 & 0.0 & 0 \\
\hline
$K^0_S$ & 0.49761 & 0 & 0.0 & 1 & 0.0 & 0 \\
\hline
$K^+$ & 0.49368 & 1 & 0.0 & 1 & 0.0 & 0 \\
\hline
$\eta$ & 0.54786 & 0 & 0.0 & 1 & 0.0 & 0 \\
\hline
$\eta'$ & 0.95778 & 0 & 0.0 & 1 & 0.0 & 0 \\
\hline
$\rho^+$ & 0.77526 & 1 & 1.0 & 3 & 1.60 & 0 \\
\hline
$\rho^0$ & 0.77526 & 0 & 1.0 & 3 & 0.0 & 0 \\
\hline
$\omega$ & 0.78265 & 0 & 1.0 & 3 & 0.0 & 0 \\
\hline
$\phi$ & 1.01946 & 0 & 1.0 & 3 & 0.0 & 0 \\
\hline
$K^{*+}$ & 0.89166 & 1 & 1.0 & 3 & 2.3 & 0 \\
\hline
$K^{*0}$ & 0.89166 & 0 & 1.0 & 3 & -0.183 & 0 \\
\hline
$p$ & 0.93827 & 1 & 0.5 & 2 & 5.5857 & 1 \\
\hline
$n$ & 0.93957 & 0 & 0.5 & 2 & -3.8263 & 1 \\
\hline
$\Lambda$ & 1.11568 & 0 & 0.5 & 2 & -1.226 & 1 \\
\hline
$\Sigma^+$ & 1.18937 & 1 & 0.5 & 2 & 6.232 & 1 \\
\hline
$\Sigma^0$ & 1.19264 & 0 & 0.5 & 2 & 2.011 & 1 \\
\hline
$\Sigma^-$ & 1.19745 & -1 & 0.5 & 2 & -2.96 & 1 \\
\hline
$\Xi^0$ & 1.31486 & 0 & 0.5 & 2 & -1.834 & 1 \\
\hline
$\Xi^-$ & 1.32171 & -1 & 0.5 & 2 & -1.834 & 1 \\
\hline
$\Delta^{++}$ & 1.232 & 2 & 1.5 & 4 & 5.37 & 1 \\
\hline
$\Delta^+$ & 1.232 & 1 & 1.5 & 4 & 2.27 & 1 \\
\hline
$\Delta^0$ & 1.232 & 0 & 1.5 & 4 & 0.017 & 1 \\
\hline
$\Delta^-$ & 1.232 & -1 & 1.5 & 4 & -2.17 & 1 \\
\hline
$\Sigma^{*+}$ & 1.3837 & 1 & 1.5 & 4 & 2.50 & 1 \\
\hline
$\Sigma^{*0}$ & 1.3837 & 0 & 1.5 & 4 & 0.26 & 1 \\
\hline
$\Sigma^{*-}$ & 1.3837 & -1 & 1.5 & 4 & -1.99 & 1 \\
\hline
$\Xi^{*0}$ & 1.532 & 0 & 1.5 & 4 & -1.82 & 1 \\
\hline
$\Xi^{*-}$ & 1.535 & -1 & 1.5 & 4 & -1.82 & 1 \\
\hline
$\Omega^-$ & 1.67245 & 0 & 1.5 & 4 & -2.40 & 1 \\
\hline
$N(1440)^+$ & 1.440 & 1 & 0.5 & 2 & 1.47 & 1 \\
\hline
$N(1440)^0$ & 1.440 & 0 & 0.5 & 2 & 0.84 & 1 \\
\hline
$\Lambda(1405)$ & 1.405 & 0 & 0.5 & 2 & -0.6 & 1 \\
\hline
$\Lambda(1520)$ & 1.520 & 0 & 1.5 & 2 & -1.0 & 1 \\
\hline
$N(1535)^+$ & 1.535 & 1 & 0.5 & 2 & 0.0 & 1 \\
\hline
$N(1535)^0$ & 1.535 & 0 & 0.5 & 2 & 0.0 & 1 \\
\hline
$N(1650)^+$ & 1.650 & 1 & 0.5 & 2 & 0.0 & 1 \\
\hline
$N(1650)^0$ & 1.650 & 0 & 0.5 & 2 & 0.0 & 1 \\
\hline
$N(1720)^+$ & 1.720 & 1 & 1.5 & 4 & 0.0 & 1 \\
\hline
$N(1720)^0$ & 1.720 & 0 & 1.5 & 4 & 0.0 & 1 \\
\hline
$\Delta(1700)^{++}$ & 1.700 & 2 & 1.5 & 4 & 0.0 & 1 \\
\hline
$\Delta(1700)^+$ & 1.700 & 1 & 1.5 & 4 & 0.0 & 1 \\
\hline
$\Delta(1700)^0$ & 1.700 & 0 & 1.5 & 4 & 0.0 & 1 \\
\hline
$\Delta(1700)^-$ & 1.700 & -1 & 1.5 & 4 & 0.0 & 1 \\
\hline
$\Delta(1900)^{++}$ & 1.900 & 2 & 1.5 & 4 & 0.0 & 1 \\
\hline
$\Delta(1900)^+$ & 1.900 & 1 & 1.5 & 4 & 0.0 & 1 \\
\hline
$\Delta(1900)^0$ & 1.900 & 0 & 1.5 & 4 & 0.0 & 1 \\
\hline
$\Delta(1900)^-$ & 1.900 & -1 & 1.5 & 4 & 0.0 & 1 \\
\hline
$\Lambda(1670)$ & 1.670 & 0 & 1.5 & 4 & 0.0 & 1 \\
\hline
$\Lambda(1800)$ & 1.800 & 0 & 1.5 & 4 & 0.0 & 1 \\
\hline
$\Sigma(1750)^+$ & 1.750 & 1 & 1.5 & 4 & 0.0 & 1 \\
\hline
$\Sigma(1750)^0$ & 1.750 & 0 & 1.5 & 4 & 0.0 & 1 \\
\hline
$\Sigma(1750)^-$ & 1.750 & -1 & 1.5 & 4 & 0.0 & 1 \\
\hline
$\Xi(1820)^0$ & 1.820 & 0 & 1.5 & 4 & 0.0 & 1 \\
\hline
$\Xi(1820)^-$ & 1.820 & -1 & 1.5 & 4 & 0.0 & 1 \\
\hline
\end{tabular}
\end{table*}

\bibliographystyle{apsrev4-2}
{}

\end{document}